\documentclass[letterpaper,10pt,conference]{ieeeconf}

\IEEEoverridecommandlockouts
\renewcommand{\baselinestretch}{1}

\usepackage{amsmath}
\usepackage{amssymb}%
\usepackage{relsize}
\usepackage{amsfonts}%
\usepackage{mathtools}
\usepackage{graphicx}
\usepackage[labelformat=simple]{subcaption}
\usepackage{epstopdf}
\graphicspath{{./epsfiles/}} 
\usepackage{stackengine}
\usepackage{algorithm,algorithmicx,algpseudocode}
\usepackage[margin=1in]{geometry}
\usepackage{color}
\usepackage{filecontents}
\usepackage{flushend}

\newcommand{\real}{\mathbb{R}}

\newcommand{\Gcal}{\mathcal{G}}
\newcommand{\G}{\Gcal}
\newcommand{\V}{\mathcal{V}}
\newcommand{\E}{\mathcal{E}}
\newcommand{\N}{\mathcal{N}}
\renewcommand{\S}{\mathcal{S}}
\newcommand{\U}{\mathcal{U}}
\newcommand{\Vprod}{\V_{1 \times 2}}

\newcommand{\alg}{\texttt{SPECTRE}}
\newcommand{\algone}{\texttt{EstimateSeeds}}
\newcommand{\algtwo}{\texttt{SafeExpand}}
\newcommand{\algthree}{\texttt{LooseExpand}}

\newcommand{\M}{\mathcal{M}}

\newcommand{\1}{\mathbf{1}}
\algnewcommand\algorithmicinput{\textbf{Input:}}
\algnewcommand\algorithmicoutput{\textbf{Output:}}
\algnewcommand\Input{\item[\algorithmicinput]}%
\algnewcommand\Output{\item[\algorithmicoutput]}%

\allowdisplaybreaks

\begin{document}
\title{SPECTRE: Seedless Network Alignment via Spectral Centralities}
\author{Mikhail Hayhoe, Jorge Barreras, Hamed Hassani, Victor M. Preciado \thanks{\noindent The authors are with the Department of Electrical \& Systems Engineering at the University of Pennsylvania. \texttt{mhayhoe}@seas.upenn.edu.}
}

\maketitle

\begin{abstract}
Network alignment consists of finding a structure-preserving correspondence between the nodes of two correlated, but not necessarily identical, networks. This problem finds applications in a wide variety of fields, from the alignment of proteins in computational biology, to the de-anonymization of social networks, as well as recognition tasks in computer vision.

In this work we introduce $\alg$, a scalable algorithm that uses spectral centrality measures and percolation techniques. Unlike most network alignment algorithms, $\alg$ requires no seeds (i.e., pairs of nodes identified beforehand), which in many cases are expensive, or impossible, to obtain. Instead, $\alg$ generates an initial noisy seed set via spectral centrality measures which is then used to robustly grow a network alignment via bootstrap percolation techniques. We show that, while this seed set may contain a majority of incorrect pairs, $\alg$ is still able to obtain a high-quality alignment. Through extensive numerical simulations, we show that $\alg$ allows for fast run times and high accuracy on large synthetic and real-world networks, even those which do not exhibit a high correlation.
\end{abstract}

\section{Introduction}
\begin{figure}[!b]
  \centering
  \includegraphics[width=0.7\linewidth]{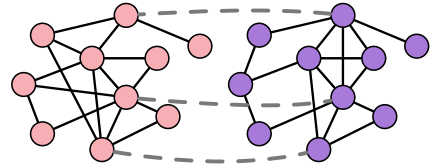}
  \caption{Correct correspondence of some nodes in two correlated networks, illustrated by dashed lines. A network alignment algorithm should find these correspondences.}\label{fig:alignment_ex}
\end{figure}

\textit{Network alignment} consists of finding a structure-preserving correspondence between the nodes of two correlated, not necessarily identical, networks. An accurate solution to this problem would address central issues in different fields, varying from the deanonymization of social networks, to recognition tasks in computer vision, to the alignment of proteins in computational biology. An example of two correlated networks is given in Figure~\ref{fig:alignment_ex}, with a correspondence that is indicated by the layout of the nodes.

An application of network alignment can be found in social network analysis, where it is possible to discover the identities of  individuals in an anonymous network by aligning its structure with that of a correlated network in which nodes are identified~\cite{AN-VS:09}. From a marketing perspective, finding individuals which play similar roles across platforms allows advertisers to integrate information from different domains in order to target ads and product recommendations~\cite{YZ:14}. In computer vision, network alignment is used for tasks such as object recognition~\cite{AB-TB:05}, image registration~\cite{ML-MH:05}, or symmetry analysis~\cite{JH-ML:06}. In these problems, nodes may represent salient points, lines, shapes, or other features in images and edges are used to encode distances between them. A further application can be found in computational biology; in particular, the study of protein-protein interaction (PPI) networks~\cite{VS-TM:14,EK-HH:16,NMD-KB-NP:17}. PPI networks provide an understanding of the system-level functions of each protein, as well as insights into how biological motifs are conserved through evolution. However, due to mutations, these corresponding proteins often have different compositions~\cite{Sharan2005} and, thus, the alignment of PPI networks needs to rely on their structural correlation. Another biological application of network alignment can be found in the problem of determining gene-disease causation~\cite{NJ-HT:14}, where the alignment of disease and PPI networks can be used to produce high-quality gene-disease candidates.

In a pioneering work, Narayanan and Shmatikov~\cite{AN-VS:09} succeeded in de-anonymizing a large-scale dataset from Netflix using publicly available auxiliary information on some users, which sparked controversy and contributed to a data privacy lawsuit~\cite{KZ:12}. Subsequent papers on the topic of network alignment assumed the availability of side information in the form of a seed set, i.e., a set of correctly-aligned nodes. Such seed set might be used in a \textit{seed-and-expand} strategy in which percolation techniques are used to ``grow'' a correct alignment throughout the nodes ~\cite{LY-MG:13}. Alternatively, the alignment can be grown locally by using the Hungarian algorithm ~\cite{OK-NP:11} or other relaxations of quadratic optimization problems ~\cite{RP-CK:12}. The inclusion of prior information in the form of a seed set typically results in much higher precision (see, e.g., FINAL~\cite{SZ-HT:16}); however, in many cases such a seed set is difficult, if not impossible, to obtain. 

Another strategy in the literature involves constructing a node-specific signature which is then used to align the networks. This is done by matching pairs of nodes with similar signatures. Many such signatures have been proposed, ranging from simple neighborhood statistics~\cite{OK-TM:10,EM-JX:18}, to spectral signatures~\cite{SF-GQ:16,RP-CK:12}, to distance to ``important'' nodes~\cite{AY-UC:18}, to more complex networks embeddings~\cite{MH-HS:18}. However, the matching step in many of these algorithms has time complexity of $O(n^2)$ or higher, which quickly becomes

\noindent infeasible for even moderately-sized networks. Algorithms like REGAL \cite{MH-HS:18} and LowRankAlign \cite{SF-GQ:16} are exceptions, since they use low rank approximations to obtain scalable algorithms.  

The \textit{seed-and-expand} as well as the signature-based algorithms described above present a critical limitation: they only produce high quality alignments when the two networks to be aligned have very high edge correlation (e.g., see Figure 4 in~\cite{MH-HS:18}). This limitation poses a great challenge for the applicability of these algorithms, since many interesting real-world applications do not assume near-perfect correlation. This critical limitation in state-of-the-art algorithms can be intuitively understood as follows. In the seed-and-expand case, a percolation process will propagate incorrect alignments in a cascading fashion; in signature-based algorithms, it is very challenging to create node-specific signatures that are global and are robust to moderate structural correlations.
To the best of our knowledge, there are no network alignment algorithms that obtain high accuracy and scalability without assuming near-perfect correlation of the networks or an initial set of correct pairs. 

In this paper, we present $\alg$: a scalable algorithm that uses spectral centrality measures together with bootstrap percolation techniques~\cite{SJ-TL:12,EK-HH:15} to align networks with high accuracy. Unlike most network alignment algorithms, $\alg$ requires no seeds (i.e., pairs of nodes identified beforehand) or side information. Instead, the algorithm is based on a \emph{seed-and-expand} strategy; in the \emph{seed} phase, $\alg$ generates an initial noisy estimate set via spectral centrality measures which is then used in the \emph{expand} phase to robustly grow an alignment of the whole network. We show that while this seed estimate may contain a majority of incorrect pairs, this noise has little impact on the final alignment.

We present extensive numerical results describing the performance of our algorithm, including comparisons to existing algorithms in social and biological networks. As our results demonstrate, $\alg$ is able to align moderately correlated, large-scale networks with high accuracy. Moreover, $\alg$ shows a noticeable improvement over the state-of-the-art methods in aligning PPI networks. For example, on PPI networks of \textit{C. jejuni} and \textit{E. coli} bacteria, the best performance in the literature for two popular metrics, called \emph{Edge Correctness} and \emph{Induced Conserved Structure} (ICS) score~\cite{RP-CK:12}, are $24\%$ and $9\%$, respectively. However, by using $\alg$, we obtain a $32\%$ edge correctness and a $35 \%$ ICS score.

Our contributions can be summarized as follows:
\begin{itemize}
\item We introduce a new algorithm, based on iterated bootstrap percolation, called $\alg$. To the best of our knowledge, this is the first scalable algorithm able to accurately and robustly align pairs of networks exhibiting moderate correlation using no prior information.
\item We propose a method for generating an initial seed set estimate using eigenvector centrality to rank nodes. This noisy seed set may have a majority of incorrect pairs; however, $\alg$ can successfully screen these incorrect pairs using bootstrap percolation.
\item Through extensive numerical experiments, we show that $\alg$ can recover high-accuracy alignments on both synthetic and real-world networks. Moreover, we compare $\alg$ to other algorithms in the literature, showing that it outperforms them when networks exhibit ground-truth correlation below $95\%$.
\end{itemize}

\section{Related Work}
We can broadly categorize the algorithms found in the literature in two categories: (i) \textit{seed-and-expand} type algorithms, and (ii) algorithms that match nodes according to their similarity given by some signature or embedding. While the \textit{seed-and-expand} type of algorithms exhibit high quality performance in terms of precision and scalability, they make the critical assumption that the user has a seed set of initial pairs of nodes that are correctly aligned. In most real world applications such a seed set could be very costly to obtain, if possible at all. Although \textit{signature-similarity} based algorithms can potentially overcome the need for a seed set, their performance is highly dependant on the type of node signature used and the construction of the similarity matrix is computationally demanding, rendering most of these algorithms unscalable. 

The idea of using a \textit{seed set} to align datasets can be traced back to Narayanan and Shmatikov's 2009 paper~\cite{AN-VS:09}, where the authors used side information (in the form of an attribute matrix) to de-anonymize large scale sparse datasets. In the context of networks, the work by Pedarsani and Grossglauser \cite{PP-MG:11} was the first to give a theoretical treatment to the problem of network alignment, as well as the first to introduce the $G(n,p;s)$ network generation model, which is widely used to generate correlated networks on which to test algorithms. These pioneering papers gave theoretical grounds to many other algorithms which assume side information in the form of a \textit{seed set} \cite{OK-NP:11,RS-JX:07,LY-MG:13}. In particular, in \cite{LY-MG:13} the authors introduce an algorithm that uses ideas from \textit{bootstrap percolation} \cite{MA-JL:88} namely, starting from a \textit{seed set}, additional pairs are aligned if there are at least $r$ aligned pairs that are ``neighbors'' of it (a precise definition of ``neighboring pairs'' will be provided later). Bootstrap percolation methods are both scalable and accurate, and some variations of them, for example ~\cite{EK-HH:15}, can considerably reduce the size of the \textit{seed set} required for good performance.  

Another family of algorithms attempts to solve the network alignment problem by designing node-level signatures and then aligning nodes with similar signatures. Many such signatures have been proposed, ranging from simple neighborhood statistics~\cite{OK-TM:10,EM-JX:18}, to spectral signatures~\cite{SF-GQ:16,RP-CK:12}, to distance to ``important'' nodes~\cite{AY-UC:18}, to more complex networks embeddings~\cite{MH-HS:18}. This approach has several scalability challenges since constructing a full similarity matrix for the nodes and obtaining a maximum weight matching cannot be solved (exactly) in linear time. Such computational considerations have motivated the development of multiple algorithms in the literature trying to combine a rich node signature with fast approximation algorithms for node matching, often as separate components. For example, the GRAAL family of algorithms uses a signature based on graphlet-degree distributions and matches nodes with a range of methods ranging from the Hungarian algorithm~\cite{OK-TM:10} to \textit{seed and expand}~\cite{OK-NP:11} methods. Some of the most notable recent developments are signature-based algorithms like REGAL \cite{MH-HS:18}, FINAL \cite{SZ-HT:16} and gsaNA \cite{AY-UC:18} which, making use of low-rank approximations and dimensionality reduction techniques, scale well to networks of hundreds of thousands of nodes. However, it is empirically observed that these algorithms only produce high-quality alignments when the two networks have near-perfect correlation. To the best of our knowledge, there is still a need for a robust, scalable and \textit{seedless} algorithm that produces high-quality alignments even on moderately correlated networks.

A critical part of the literature in network alignment deals with understanding and curbing error propagation in \textit{seed-and-expand} algorithms~\cite{LY-MG:13,EK-HH:15}. Most notably, the authors of \cite{EK-HH:15} describe a bootstrap percolation strategy that is robust to the presence of incorrect pairs in the seed set and provably percolates to the whole network (on certain synthetic graphs). Simply put, this algorithm is more robust because it defers the matching of a pair of nodes until it accumulates enough ``neighboring candidate pairs'' (referred to as \textit{marks} in the paper). Our proposed algorithm leverages this idea, in conjunction with a boosting strategy, to overcome the dependance on a \textit{seed set}, allowing us to obtain high-quality alignments even in moderately correlated networks. 

In recent years, we find a growing literature regarding attributed network alignment and multiple network alignment. Most notably, the recent paper by Kazemi and Grosglausser~\cite{EK-MG:18} proposes the creation of a seed set in combination with a seed-and-expand strategy in the context of aligning multiple attributed networks. In contrast with the work in \cite{EK-MG:18}, the aim of our work is improving the performance of state-of-the-art algorithms for the purely structural alignment problem.

\section{Preliminaries}\label{sec:prelim}
\begin{table}[ht]
\caption{Notation}
\begin{tabular}{r l}
Symbol & Description\\
\hline
$\Gcal$ 				& Graph (undirected) \\
$\Gcal_1,~\Gcal_2$		& Correlated graphs \\
$\Gcal_{1\times 2}$		& Product graph \\
$\V$					& Vertex set, i.e., $ \{1,\ldots,n\}$ \\
$\E$					& Edge Set, subset of $\V \times \V$ \\
$i \sim j$				& $\{i,j\} \in \E$ \\
$\N_i(\Gcal) $	& Neighbors of node $i$ in graph $\Gcal$ \\
$\N_{(i,j)}$	& Neighbors of pair $(i,j)$ in product graph $\G_{1 \times 2}$ \\
$D(\Gcal)$	& Degree matrix, i.e., $\text{diag}\{|\N_1(\Gcal)|,\ldots,|\N_n(\Gcal)|\}$ \\
$A = A(\Gcal)$			& Adjacency matrix, $ [A]_{ij} = \1\{i\sim j\} $ \\
$\lambda_i(M)$			& $i$th eigenvalue of $M \in \real^{n \times n}$ (decreasing magn.) \\
$\lambda_{max}(M)$		& largest eigenvalue of $M$, i.e., $\lambda_1(M)$ \\
$s(i,j)$				& score of pair $(i,j)$ from $\G_{1 \times 2}$
\end{tabular}
\end{table}

In this work we consider undirected graphs\footnote{We use graph and network interchangeably.} $ \Gcal = (\V,\E) $, where $ \V $ is the set of nodes and $ \E $ is the set of unweighted edges. We assume $\Gcal$ is simple, i.e., it has no self-loops or multi-edges. 
The product graph $\Gcal_{1\times 2}$ of two networks $\Gcal_1$ and $\Gcal_2$ is the graph with vertex set $\V_{1\times 2} = \V_1 \times \V_2$, and edge set $\big\{\{(i,j),(u,v)\} :  \{i,u\} \in \E_1, \{j,v\} \in \E_2\big\} \subseteq \V_{1\times 2} \times \V_{1\times 2}$. The set of all neighbouring pairs of $(i,j)$ in the product graph $\G_{1 \times 2}$ are denoted by $\N_{(i,j)} = \big\{(u,v)\in\V_{1\times 2} : \{i,u\} \in \E_1, \{j,v\} \in \E_2\big\}$.


Formally, the problem of network alignment on two graphs $\Gcal_1 = (\V_1,\E_1)$ and $\Gcal_2 = (\V_2,\E_2)$ is to find a matching set $\M \subset \V_1 \times \V_2$, so that $(i,j) \in \M$ means $i \in \V_1$ corresponds to the same unique entity as $j \in \V_2$, which we write as $i \leftrightarrow j$. For example, in the context of social networks, we may imagine matching the account of an individual on Twitter to a Facebook account owned by the same individual. However, since the sets of users may be different in both networks, we may only hope to match pairs of nodes in $\V_1 \cap \V_2$. Given a matching $\M$, we will define the correspondence $f_{\M}: \V_1 \to \V_2$ as $f_{\M}(i) = j$ if $(i,j) \in \M$, and undefined otherwise. Similarly, we define the set of matched nodes in $\V_2$ as $\M(\V_1) = \{f_{\M}(i) : i \in \V_1,\text{ $i$ matched}\}$ and the set of matched edges as $\M(\E_1) = \{(f_{\M}(i), f_{\M}(j)) : (i,j) \in \E_1; \text{ $i,j$ matched}\}$, which may include edges not present in $\E_2$.

In order to measure the performance of our proposed algorithm, $\alg$, it is necessary to have correlated networks where the true matching of nodes is available. In our numerical experiments, we create such networks from real-world data~\cite{BR-RD-RS-CS:18}. In order to symmetrically generate two networks while preserving access to the the true matching information, we use the following procedure. We start with an arbitrary graph $\Gcal = (\V,\E)$ and  generate a random graph $\tilde{\Gcal}_1$ by independently subsampling each edge in $\Gcal$ with a probability $1 - s$. Hence, $\tilde{G}_1$ has the same node set as $\Gcal$ and an expected number of edges $(1-s)|\E|$. We repeat this procedure, independently, to obtain a second graph $\tilde{\Gcal}_2$. Notice that, as a result of subsampling edges, $\tilde{\Gcal}_1$ and/or $\tilde{\Gcal}_2$ may become disconnected. To overcome this issue, we follow the iterative procedure described below. In a first step, we find the largest connected components of the two graphs, denoted by $\tilde{C}_1$ and $\tilde{C}_2$ respectively. We then look at the subgraphs of $\tilde{\Gcal}_1$ and $\tilde{\Gcal}_2$ induced by the nodes in $\tilde{C}_1 \cap \tilde{C}_2$. If these induced subgraphs are connected, we call them $\Gcal_1$ and $\Gcal_2$ and take them as the pair of networks to be aligned; if they are disconnected, we find their largest connected components and repeat these steps. This approach to generate correlated random graphs was introduced in~\cite{PP-MG:11} where the following edge similarity measure was also proposed:
\begin{align*}
	Sim_e(\Gcal_1,\Gcal_2) = 2\sum_{i,j\in \V_1}\frac{\1\big\{\{i,j\} \in \E_1, \{i,j\} \in \E_2\big\}}{|\E_1| + |\E_2|},
\end{align*}
although there are other ways to measure the similarity of two graphs~\cite{LZ-GV:08}.

\section{Algorithms}\label{sec:alg}
In this section we introduce $\alg$, a scalable algorithm able to solve the network alignment problem with high accuracy in the absence of side information.
$\alg$ uses spectral properties of $\Gcal_1$ and $\Gcal_2$ to create a noisy initial seed set $\S$,  which will contain a number of correct pairs and many incorrect ones, as we will describe in Subsection~\ref{subsec:algone}. This initial set is not a proper matching, since the same node can be present in numerous pairs.  $\S$ is then used to build a confident seed estimate $\M_0$, where nodes appear in at most one pair, following a strict percolation procedure described in Subsection~\ref{subsec:algtwo}. $\alg$ then performs a backtracking step, resetting the matching and using $\M_0$ as a new seed estimate. In Subsection~\ref{subsec:algthree}, we propose a relaxed, looser, percolation which uses the confident seed estimate to percolate a matching $\M$ over the networks. Finally, if the percolation does not grow above a fraction $f$ of the networks' size, we backtrack by using the final matching as input to the algorithm again as if it were a noisy seed set, and the process is repeated. Through this backtracking procedure $\alg$ is able to build a final matching that has significantly higher accuracy, even when the networks exhibit low correlation. Typically we choose $f = 3/4$, but this parameter may be increased if a larger matching is desired. In Algorithm~\ref{alg} below, we provide the general structure of $\alg$, and in the following subsections we describe each subroutine in detail.


\begin{algorithm}[ht]
	\begin{algorithmic}
    \Input{$\Gcal_1$,$\Gcal_2$ are graphs to align; $k$ is number of top seeds; $w$ is size of window; $r$ is token threshold}
        \Output{$\M$ is the matching}
    	\State $C_1, C_2 \gets $ eigenvector centralities of $\Gcal_1$ and $\Gcal_2$ (resp.)
        \State $ \S \gets $ \algone($k,w,C_1,C_2$)
		\While{$ | \M | < f*\min\{|\V_1|,|\V_2|\} $}
		\State $ \M_0 \gets $ \algtwo($\S,r$)		\Comment{Confident perc.}
		\State $ \S \gets \M_0 $					\Comment{Backtracking}
		\State $ \M \gets $ \algthree($\S$)		\Comment{Relaxed perc.}
        		\State $ \S \gets \M $					\Comment{Backtracking}
        \EndWhile
		\State \Return $ \M $
	\caption{$\alg(\Gcal_1,\Gcal_2,k,w,r)$}\label{alg} 
	\end{algorithmic}
\end{algorithm}

\subsection{$\algone$ subroutine}\label{subsec:algone}

\algone, the first subroutine in \alg, constructs a noisy seed set estimate which should contain some number of correct pairs, i.e. pairs of nodes that are correctly matched across the networks. To ensure that this occurs, nodes across networks should be matched using a procedure that is robust to perturbations in the network structure. In $\alg$, this procedure is based on comparing the spectral centralities of nodes in different networks; an in-depth description of this choice is presented in Section~\ref{subsec:centrality}. In particular, in order to create a noisy seed set estimate $\S$, we rank the nodes of $\Gcal_1$ and $\Gcal_2$ by their centrality scores and keep the top $k$ most central nodes in each network.
The rationale behind this choice of potential matches is that, for correlated graphs $\Gcal_1$ and $\Gcal_2$, nodes with high centrality in $\Gcal_1$ are likely to be aligned with nodes of high centrality in $\Gcal_2$. Furthermore, the centrality ranking of matched nodes should be similar for the most central nodes. As a result, $\S$ contains $(2w + 1)k - w(w+1)$ pairs, of which no more than $ k $ represent correct matches. While higher values of $w$ increase the probability of finding $k$ correct pairs, it also increases the number of incorrect pairs by $O(k)$, and thus we must be conservative with our choice of both parameters. Taking $k = O(\log n)$ and $ w = 1,2$ performs well in practice; typically we set $k = 10\log n$ and $w = 1$.

It is noteworthy to emphasize that the noisy seed set estimate $\S$, generated by $\algone$, typically contains a large fraction of incorrect pairs. However, since correct pairs increment scores of other correct pairs more effectively than incorrect pairs may increment scores among themselves, the algorithms $\algtwo$ and $\algthree$ are able to robustly percolate and find a moderately accurate alignment, even in the presence of many incorrect pairs in $\S$. In some cases this initial iteration will in fact yield a highly accurate alignment, especially if the original network is dense.  However, when the initial iteration is not enough, the backtracking step of $\alg$ is able to boost the moderately successful alignment in order to significantly increase both the size and accuracy of the final matching. As a result, provided that there are a sufficient amount of correct pairs in the noisy seed set estimate $\S$, then $\algtwo$ and $\algthree$ will be able to overcome the presence of wrong pairs and percolate over the set of all the correct pairs.

\begin{figure}
\centering
\includegraphics[width=0.85\linewidth]{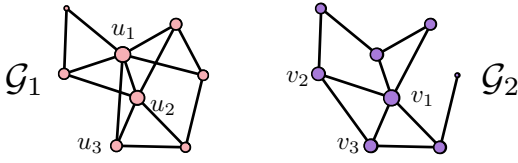}
\caption{Example of $\algone$. The size of each node denotes its relative spectral centrality score in the network, and the ground-truth correspondence is illustrated by the position of the nodes. Here with $k = 3$ and $w = 1$, $u_1$ is matched with $v_1$ and $v_2$; $u_2$ is matched with $v_1$, $v_2$ and $v_3$; and $u_3$ is matched with $v_2$ and $v_3$. Of these pairs, the seeds $(u_2,v_1)$ and $(u_3,v_3)$ are correct.}
\label{fig:ex1}
\end{figure}


\begin{algorithm}[ht]\caption{$\algone(k,w,C_1,C_2)$}
	\begin{algorithmic}
    \Input{$k$ is number of top seeds; $w$ is size of window; $C_1, C_2$ are centrality scores}
    \Output{$\S$ is the noisy seed set}
    \State $ \S \gets \emptyset $
        \For{each of top $ k $ nodes $ i \in \V_1 $ according to $ C_1 $}
		\State add pairs $(i,j)$ to $ \S $ by selecting the correspondingly
		\State ranked node $ j \in \V_2 $ according to $ C_2 $, as well as $ w $ nodes
		\State before and after $ j $ in the ranking.
		\EndFor
        \State\Return $ \S $
    \end{algorithmic}
\end{algorithm}

\subsection{Centrality}\label{subsec:centrality}



As mentioned previously, there are many works which explore the creation of node signatures for use in network alignment problems~\cite{OK-TM:10,SF-GQ:16,RP-CK:12,MH-HS:18}. Of particular interest are those signatures which are robust to perturbations in network topology. Specifically, we are interested in signatures which do not exhibit large changes in the relative rankings of the highest-scored nodes in the network when the nodes or edges are altered. For this reason, we choose a notion of \emph{centrality} as such a nodal feature. Node centralities are commonly used to measure the importance of nodes, and can be used to estimate the influence of individuals in social networks~\cite{SW-KF:94}, the importance of web pages~\cite{SB-LP:98}, or the certainty of node measurements~\cite{IP-GY:16}. In $\alg$, we use the \emph{eigenvector} centrality $C_{ev}$\footnote{Other centrality measures were tested, including PageRank, degree, betweenness, and closeness, but eigenvector performed best empirically.}, which is formally defined as
 \begin{align}
 C_{ev}(i) &= [v_1]_i, 
 \end{align}
where $v_1$ is the eigenvector of $A(\Gcal)$ for $\lambda_{max}(A)$.
%

This centrality measure is capable of being computed efficiently, even for large-scale networks. Indeed, modern algorithms allow the calculation in $O(m)$ time and storage, where $m = |\E|$, and the constants depend on $\lambda_1(A)$ and $\lambda_2(A)$~\cite{LT-DB:97}. Interestingly, in practice, perturbations of the network topology do not dramatically change the ranking induced by this centrality measure, at least for the nodes with the highest centrality.

\subsection{$\algtwo$ subroutine}\label{subsec:algtwo}

The subroutine $\algtwo$ uses the noisy seed set estimate $\S$ from $\algone$ to construct a confident seed estimate $\M_0$, which is a matching where each node is present in at most one pair (see Algorithm \ref{algtwo}). For each possible pair $(i,j)$ in $\Vprod$, $\algtwo$ builds a confidence score $s(i,j)$ by allowing other pairs to increment these scores through the edges of the product graph $\Gcal_{1\times 2}$. In practice, many of these scores will remain at zero since the networks we consider are not fully connected. As $\algtwo$ begins, each pair of nodes $(i,j)\in \S$ increases the score of all of its neighboring pairs in $\Gcal_{1\times 2}$, i.e., all pairs in $\N_{(i,j)}$\footnote{Recall $\N_{(i,j)} = \big\{(u,v)\in\V_{1\times 2} : \{i,u\} \in \E_1, \{j,v\} \in \E_2\big\}$.}, by one. Notice that this set of neighboring pairs corresponds to all the pairs of nodes in $\N_i(\Gcal_1) \times \N_j(\Gcal_2)$. It is worth remarking that the originating pair $(i,j)$ does not increment its own score. At the end of this spreading process, only pairs which are neighbors (in the product graph) of pairs in the noisy seed set estimate $\S$ will have a positive score.

In what follows, we sequentially grow the confident seed estimate $\M_0$ according to the following procedure, which repeats as long as some pair $(i,j)$ has a score at least $r$, i.e., $s(i,j) \geq r$ for some $(i,j) \in \Vprod$. Based on the percolation bounds established in~\cite{LY-MG:13}, the value of $r$ is typically chosen to be 4. First, we find the set of all pairs in $\Vprod$ with the highest score, pick one of these pairs at random, and add it to the set $\M_0$. Next, this chosen pair increments the scores of its neighboring pairs (in the product graph) by one. According to this updated score, we pick the pair with the highest score (breaking ties at random), excluding any pair containing an already matched node (i.e., any node contained in any pair in $\M_0$). We then add the chosen pair to $\M_0$, increment the scores of its neighboring pairs, and repeat this procedure until the remaining unmatched pairs have less than $r$ tokens. At the end of this procedure, we obtain the confident seed set estimate $\M_0$, representing a matching between nodes of $\Gcal_1$ and $\Gcal_2$. This matching is, in general, not perfect, since some nodes may be left unmatched.

\begin{figure}[ht]
\centering
\begin{subfigure}[th]{0.32\linewidth}
\centering
	\includegraphics[width=0.6\linewidth]{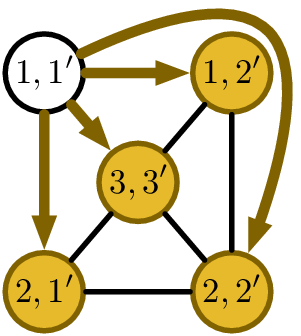}
	\caption{Seed $(1,1')$ increments scores.}
	\label{fig:build1}
\end{subfigure}
\begin{subfigure}[th]{0.32\linewidth}
\centering
	\includegraphics[width=0.6\linewidth]{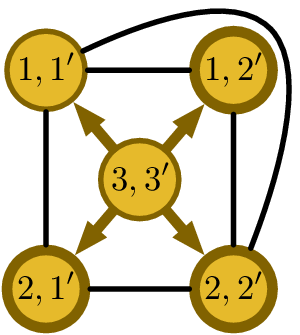}
	\caption{Seed $(3,3')$ increments scores.}
	\label{fig:build2}
\end{subfigure}
\begin{subfigure}[th]{0.32\linewidth}
\centering
	\includegraphics[width=0.6\linewidth]{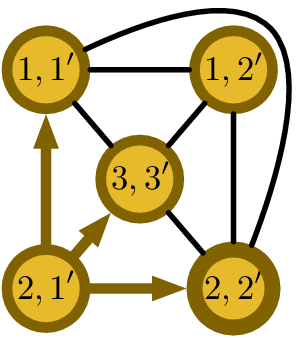}
	\caption{Seed $(2,1')$ increments scores.}
	\label{fig:build6}
\end{subfigure}
\\
\begin{subfigure}[th]{0.3\linewidth}
\centering
	\includegraphics[width=0.6\linewidth]{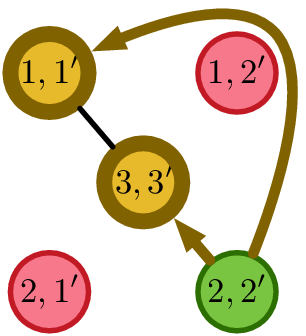}
	\caption{Pair $(2,2')$ matched and so increments scores; $(1,2')$, $(2,1')$ are removed.}
	\label{fig:build3}
\end{subfigure}
~
\begin{subfigure}[th]{0.3\linewidth}
\centering
	\includegraphics[width=0.6\linewidth]{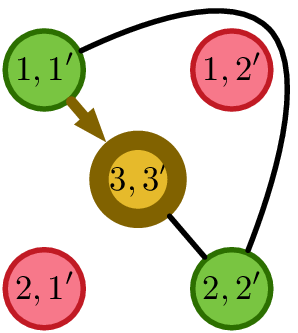}
	\caption{Break the tie at random, so $(1,1')$ matched and then increments scores.\\}
	\label{fig:build4}
\end{subfigure}
~
\begin{subfigure}[th]{0.3\linewidth}
\centering
	\includegraphics[width=0.6\linewidth]{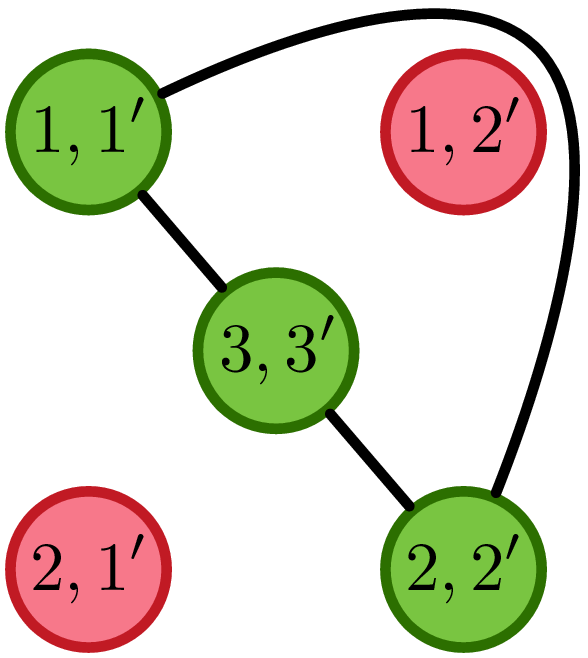}
	\caption{Match $(3,3')$. $\algtwo$ stops.\\~\\}
	\label{fig:build5}
\end{subfigure}
\caption{Example of $\algtwo$, with matching threshold $r = 3$ and noisy seed set estimate $\S = \{(1,1'),(3,3'),(2,1')\}$. White pairs have score zero, yellow have positive score (with border thickness denoting total number), green are matched, and red are removed (one or both nodes already matched in another pair). Arrows describe the direction in which scores are incremented.}
\label{fig:ex2}
\end{figure}


\begin{algorithm}[ht]\caption{$\algtwo(\S,r)$}\label{algtwo}
	\begin{algorithmic}
    \Input{$\S$ is noisy seed set; $r$ is token threshold}
    \Output{$\M_0$ is confident seed estimate}
		\State $ \M_0 \gets \emptyset $
		\State $s(u,v) \gets 0$ for ever pair $(u,v)$ \Comment{Reset scores}
		\For{each pair $(i,j) \in \S$}
			\State $s(u,v) \gets s(u,v) + 1,~\forall (u,v) \in \N_{(i,j)}$\Comment{$\uparrow$ score}
		\EndFor
		\While{any unmatched pair $(i,j)$ has $s(i,j) \geq r $ }
				\State pick $(i,j)$ randomly from highest-scoring pairs
				\State $\M_0 \gets \M_0 \cup \{(i,j)\}$\Comment{Add pair to $\M_0$}
				\State $s(u,v) \gets s(u,v) + 1,~\forall (u,v) \in \N_{(i,j)}$
		\EndWhile
        \State\Return $ \M_0 $
    \end{algorithmic}
\end{algorithm}

\subsection{$\algthree$ subroutine}\label{subsec:algthree}

The last subroutine, $\algthree$, is similar to $\algtwo$. While $\algtwo$ sets a high score threshold in order to be confident as it matches nodes, $\algthree$ is more relaxed in its acceptance of matched pairs. However, it does take into account centrality measures when breaking ties, making $\algthree$ more certain about the correctness of a pair relative to its competitors in terms of score. $\algthree$ backtracks, starting a new matching $\M$ from scratch and repeatedly growing, taking the seed set estimate $\S$ (which after the backtracking step is in fact $\M_0$) as input; see Algorithm~\ref{algthree}.

Similarly to $\algtwo$, $\algthree$ starts by having all pairs in $\S$ increase the scores of their neighboring pairs. Then, we find all the pairs composed of unmatched nodes with the highest score. Among those pairs, we select the pair with the lowest difference in the centrality measures and add it to $\M$. The selected pair then increments the score of its neighbors in the product graph by one, but only if the pair has not previously been used to increase scores. We use this procedure to iteratively add pairs to $\M$ until no pairs composed of unmatched nodes have score two or more. Then, we allow a relaxation of our percolation so our matching may spread further throughout the networks. In a rebuilding step, a new seed set $\S$ is created from scratch by taking all unmatched neighbors of matched pairs (i.e., all unmatched pairs with score exactly one), and the percolation process is repeated. This continues until no unmatched neighbors of matched pairs exist, and the final matching $\M$ is returned.

Finally, if the matching $\M$ has not grown above a fraction $f$ of the smaller network's size, we perform a backtracking step. Using the final matching $\M$ as the noisy seed set $\S$, we repeat another iteration of $\algtwo$ and $\algthree$. This boosting procedure is critical in allowing $\alg$ to perform well on networks exhibiting lower correlations, since it allows a poor-quality matching to be iteratively updated until we obtain a high-quality final matching.


\begin{figure}[!th]
\centering
\begin{subfigure}[th]{0.3\linewidth}
\centering
	\includegraphics[width=0.9\linewidth]{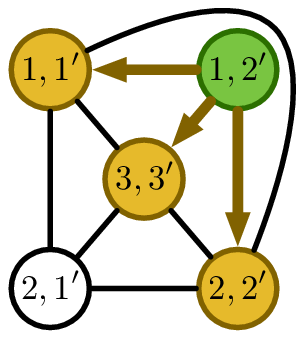}
	\caption{Pair $(1,2')$ matched, increases scores, no pairs have score $\geq 2$.}
	\label{fig:build3}
\end{subfigure}
~
\begin{subfigure}[th]{0.3\linewidth}
\centering
	\includegraphics[width=0.9\linewidth]{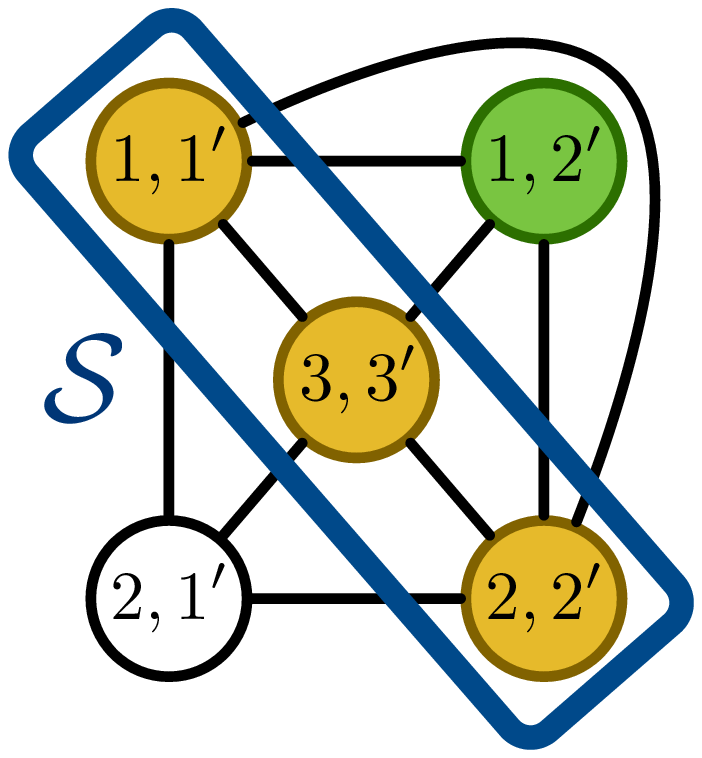}
	\caption{Rebuild set $\S$ as $\{(1,1'),(3,3'),$ $(2,2')\}$.\\}
	\label{fig:build4}
\end{subfigure}
~
\begin{subfigure}[th]{0.3\linewidth}
\centering
	\includegraphics[width=0.9\linewidth]{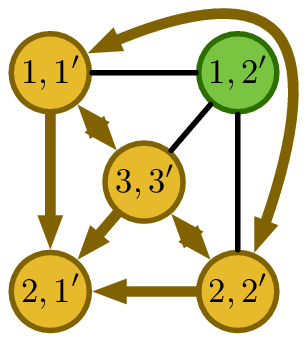}
	\caption{Pairs in $\S$ increment scores, if they have not already.}
	\label{fig:build5}
\end{subfigure}
\caption{Example of $\algthree$ rebuilding the set $\S$. We take all unmatched neighboring pairs of previously matched nodes (i.e., those pairs with score exactly one) and add them to $\S$. White pairs have score zero, yellow have positive score, and green are matched. Arrows describe the direction in which scores are incremented.}
\label{fig:ex2}
\end{figure}

\begin{algorithm}[!ht]\caption{$\algthree(\S)$}\label{algthree}
	\begin{algorithmic}
    \Input{$\S$ is confident seed estimate}
    \Output{$\M$ is the matching}
    	\State $ \M, \U \gets \emptyset $  \Comment{$ \U $ is used pairs}
	\State $s(u,v) \gets 0$ for ever pair $(u,v)$	\Comment{Reset scores}
	\While{$ | \S | > 0 $}
        	    \For{each pair $(i,j) \in \S $}
            	\State $s(u,v) \gets s(u,v) + 1,~\forall (u,v) \in \N_{(i,j)}$		
		\State $\U \gets \U \cup \{(i,j)\}$		\Comment{$(i,j)$ used to $\uparrow$ scores}
            \EndFor
			\While{any unmatched pair $(i,j)$ has $s(i,j) \geq 2 $}
				\State $\mathcal{X} \gets \arg\max s(u,v) \cap (\V_1\times\V_2\setminus\M)$
				\State $(i,j) \gets \arg\min_{(u,v) \in \mathcal{X}} | C_1(u) - C_2(v)| $
				\State $\M \gets \M \cup \{(i,j)\}$
				\If{$ (i,j) \not\in \U $}
					\State $s(u,v) \gets s(u,v) + 1,~\forall (u,v) \in \N_{(i,j)}$
					\State $\U \gets \U \cup \{(i,j)\}$
				\EndIf
			\EndWhile
			\State $ \S \gets \{(i',j') \ | \ (i',j') $  neighbor of $ (i,j) \in \M, $
			\State\qquad~~ $(i',j') \not\in \U, i'~\&~j' $ unmatched $ \}$ 
			\State\Comment{Rebuild $\S$ from scratch}
		\EndWhile
		\State \Return $ \M $
	\end{algorithmic}
\end{algorithm}

\section{Numerical Experiments}\label{sec:results}
In order to verify the effectiveness of $\alg$, we conducted extensive experiments on a variety of benchmark networks. We measure the performance of $\alg$ across four different metrics. The first is \textit{Precision}, which measures the percentage of correct pairs in the final matching $\M$. The second is \textit{Recall}, or true positive rate, which is the fraction of possible correct pairs that are identified in $\M$. Algorithms may only align a subset of the nodes, thus these two metrics evaluate different things: precision provides a notion of hit-rate on the matched nodes, while recall describes the proportion of the nodes which the algorithm was able to align. Since our algorithm may only hope to label nodes with degree at least 2 (due to how $\algthree$ matches pairs), we measure the fraction of these nodes which are in $\M$. Formally, we can define the first two metrics as follows:
\begin{align*}
	\text{Prec}(\M) &= \frac{|\{(i,j)\in\M : i \leftrightarrow j\}|}{|\M|}, \\
	\text{Recall}(\M) &= \frac{|\{(i,j)\in\M : i \leftrightarrow j\}|}{|\{v \in \V_1 \cap \V_2 : d_{1}(v) \geq 2,~d_{2}(v) \geq 2\}|},
\end{align*}
where $d_{i}(v)$ is the degree of node $v$ in graph $\G_i$. An issue with these metrics is that they require knowledge of the ground truth of the node correspondences. However, in most realistic scenarios, these correspondences would not be available. Following the approach in~\cite{RP-CK:12}, we measure the quality of our alignments using the \emph{Edge Correctness} (EC) and \emph{Induced Conserved Structure} (ICS) score. As described below, these two scores depend solely on topological information. In particular, Edge Correctness measures the fraction of matched edges from $\E_1$, denoted by $\M(\E_1)$, which are present in $\E_2$. In other words, EC measures the fraction of edges which are correctly matched by $\M$. However, EC does not penalize $\M$ for omitting edges in $\E_2$ which should be present. For this reason we also compute the ICS score, which measures the fraction of matched edges present in the subgraph of $\Gcal_2$ induced by the nodes which are matched, i.e., those nodes in $\M(\V_1)$. The ICS score penalizes the matching both for omitting edges that are present in $\E_1$ and those that are present in $\E_2$. Formally,
\begin{align*}
	\text{EC}(\Gcal_1,\Gcal_2,\M) &= \frac{|\M(\E_1)\cap\E_2|}{|\E_1|}, \\
	\text{ICS}(\Gcal_1,\Gcal_2,\M) &= \frac{|\M(\E_1)\cap\E_2|}{|\{(i,j) \in \E_2 \ : \ i,j \in \M(\V_1)\}|}.
\end{align*}

\subsection{Parameter Selection}\label{subsec:params}

\begin{figure}[!th]
	\centering
	\begin{subfigure}[th]{0.3\linewidth}
	\includegraphics[width=\linewidth]{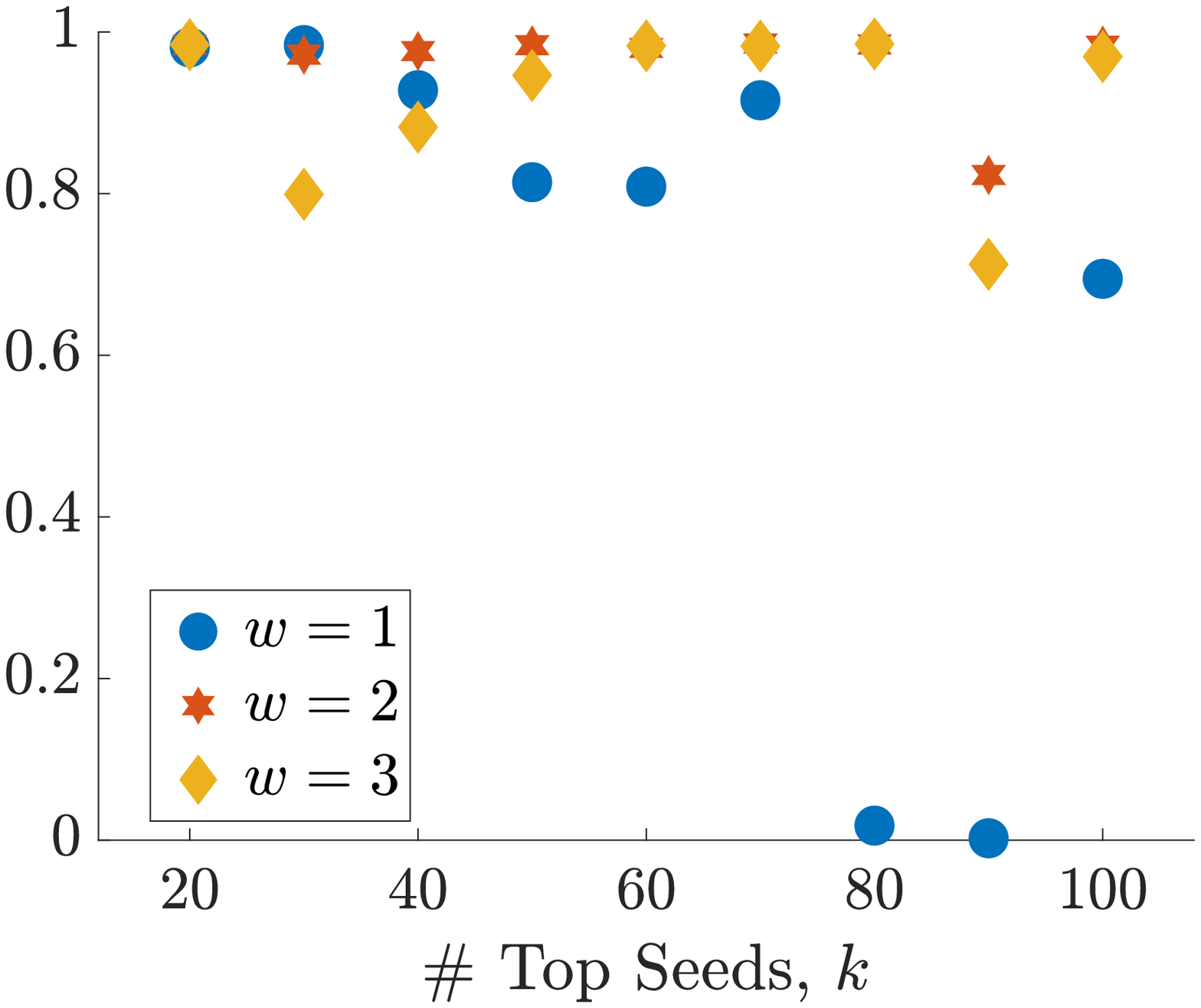}
	\caption{Precision \\~}
	\label{subfig:gem_prec}
	\end{subfigure}
	~
	\begin{subfigure}[th]{0.3\linewidth}
	\includegraphics[width=\linewidth]{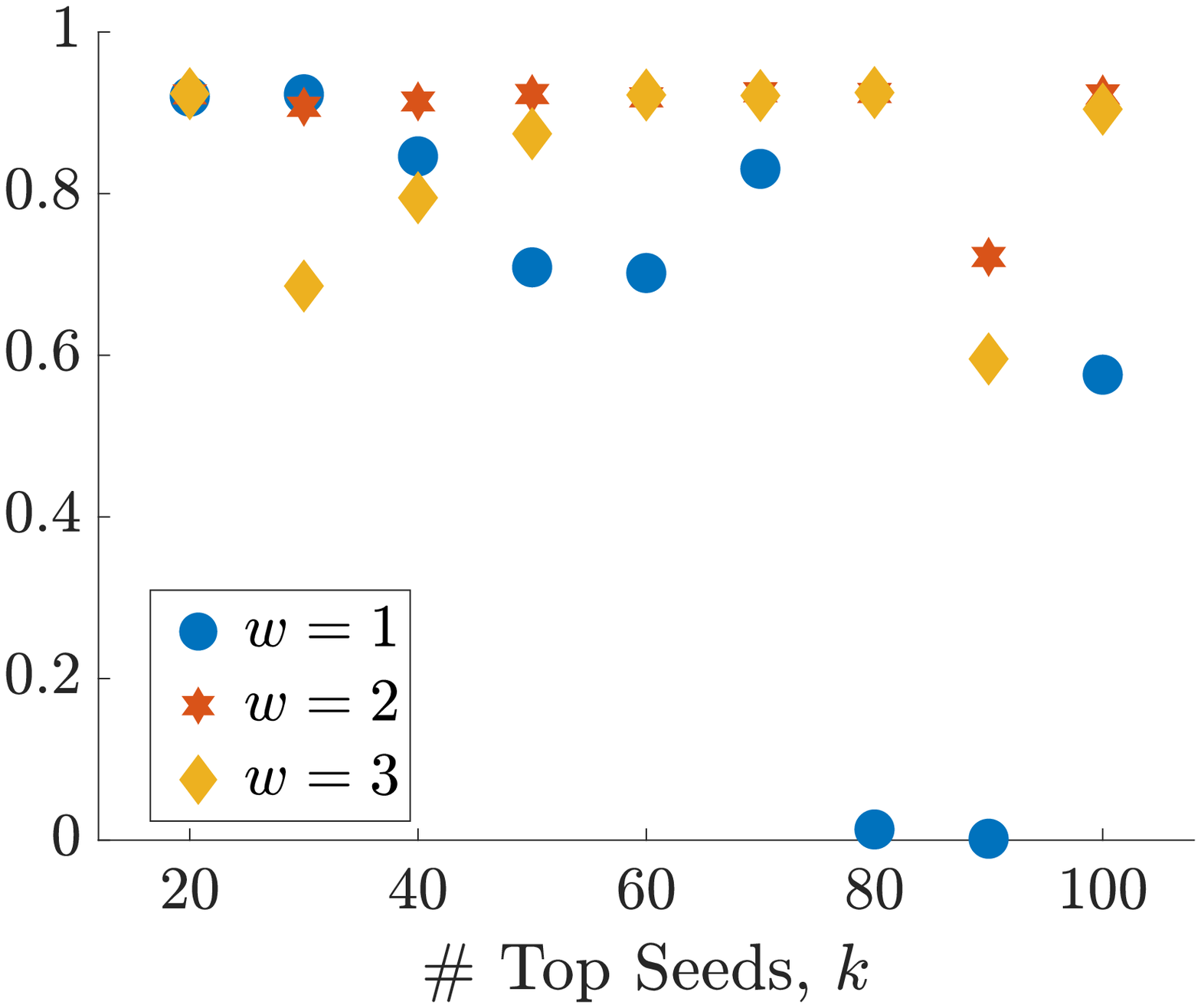}
	\caption{Recall \\~}
	\label{subfig:gem_rec}
	\end{subfigure}
	~
	\begin{subfigure}[th]{0.3\linewidth}
	\includegraphics[width=\linewidth]{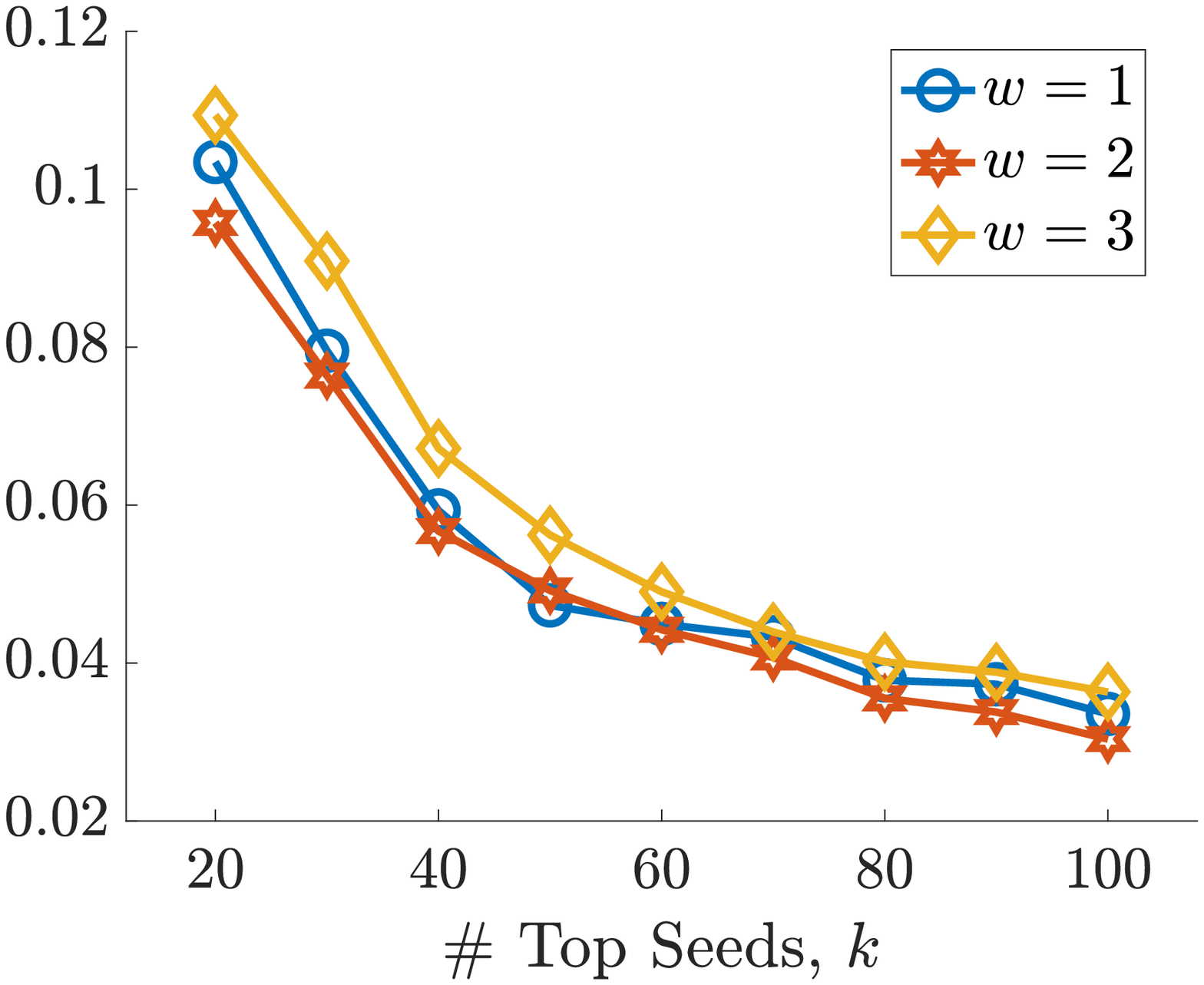}
	\caption{Percentage of correct pairs in $\S$}
	\label{subfig:gem_pct}
	\end{subfigure}
	\caption{Effect of changing $k$ and $w$ for GEMSEC-Artists network with $60\%$ correlation.}\label{fig:gem_params}
\end{figure}

\begin{figure*}[!th]
	\centering
	\begin{subfigure}{0.3\linewidth}
	\includegraphics[width=\linewidth]{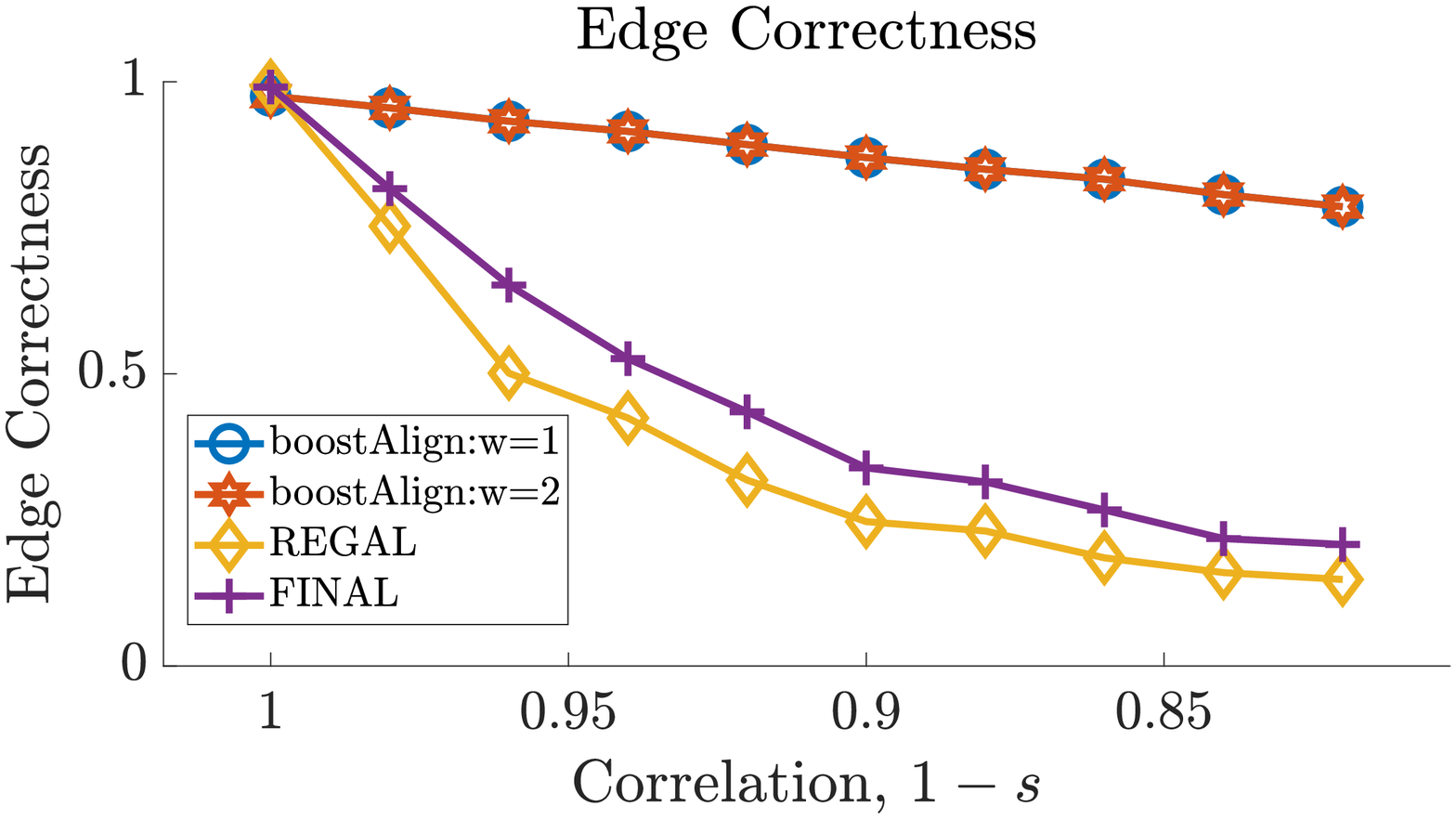}
	\label{subfig:ppi}
	\end{subfigure}
	\quad
	\begin{subfigure}{0.3\linewidth}
	\includegraphics[width=\linewidth]{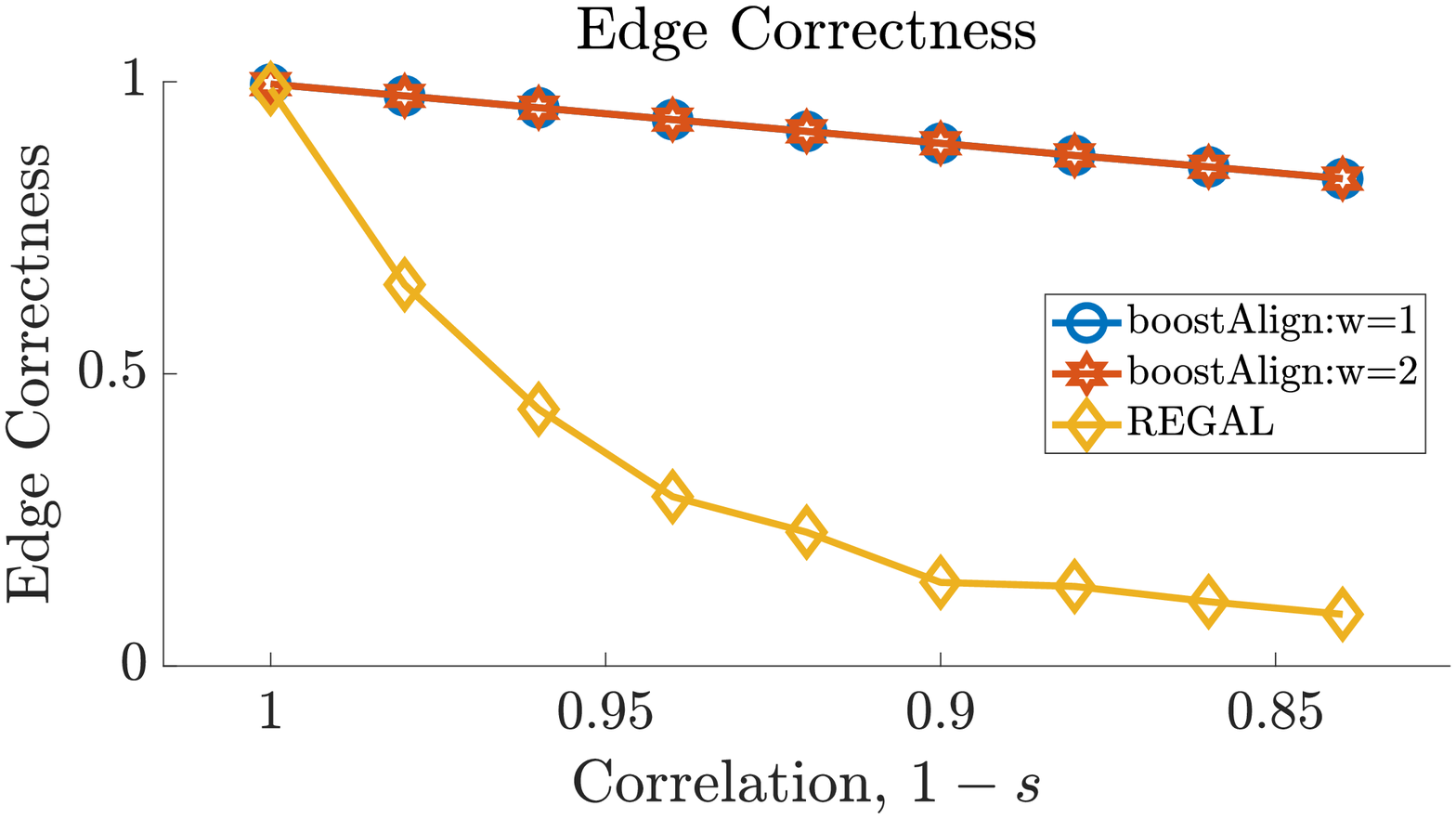}
	\label{subfig:gemsec}
	\end{subfigure}
	\quad
	\begin{subfigure}{0.3\linewidth}
	\includegraphics[width=\linewidth]{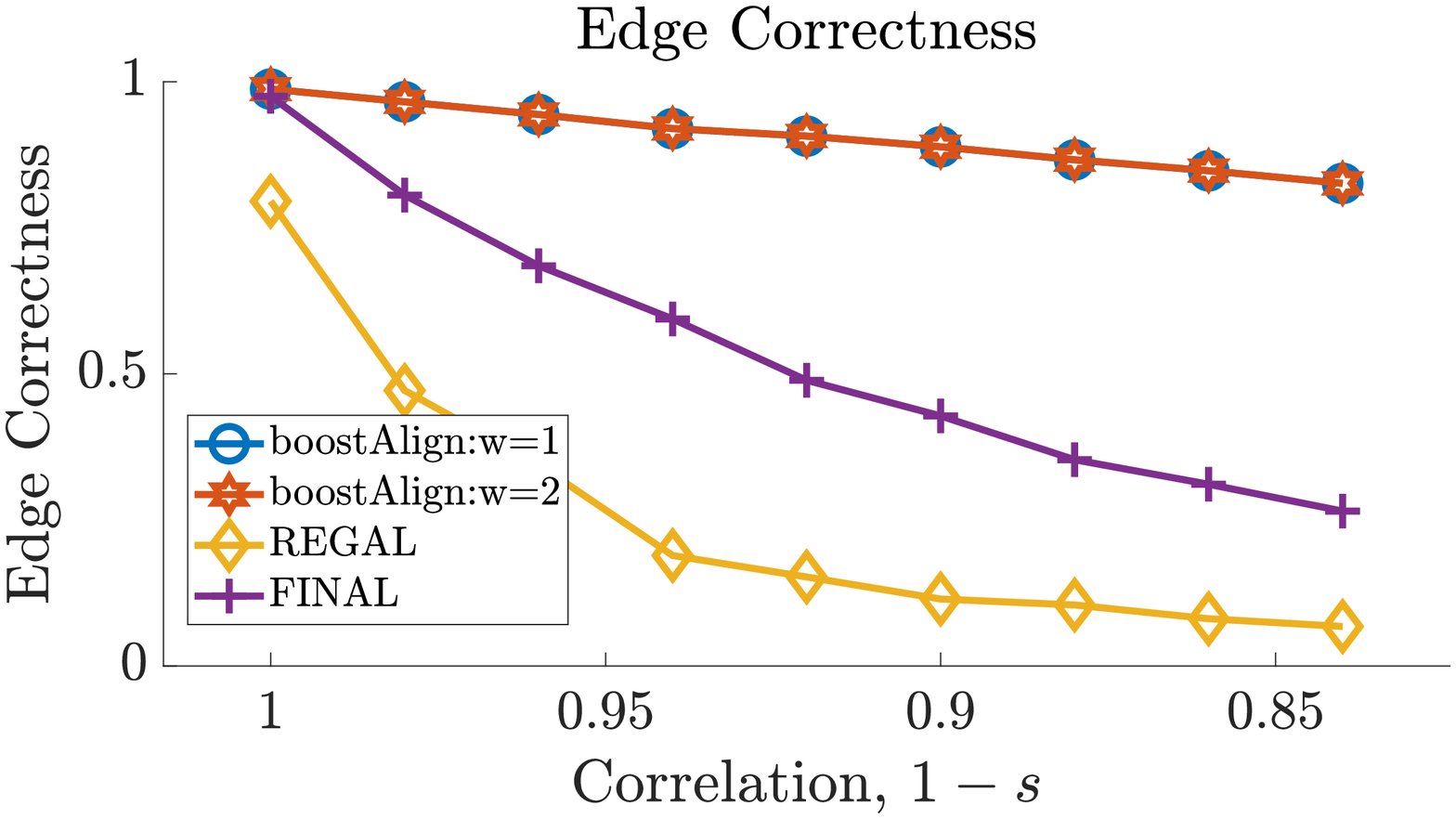}
	\label{subfig:astro}
	\end{subfigure}
	\\
	\begin{subfigure}{0.3\linewidth}
	\includegraphics[width=\linewidth]{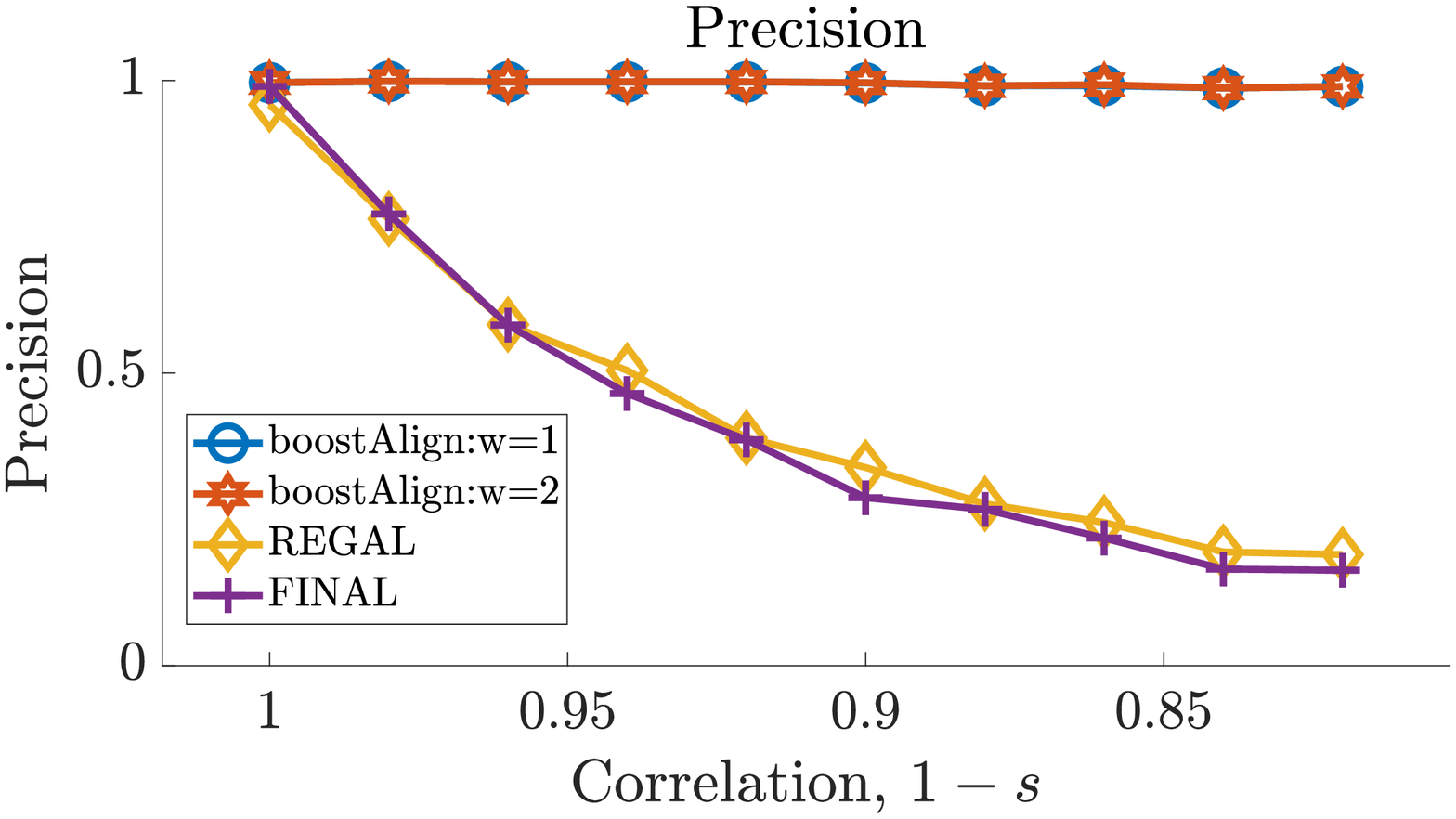}
	\caption{\textit{C. jejuni} PPI}
	\label{subfig:ppi}
	\end{subfigure}
	\quad
	\begin{subfigure}{0.3\linewidth}
	\includegraphics[width=\linewidth]{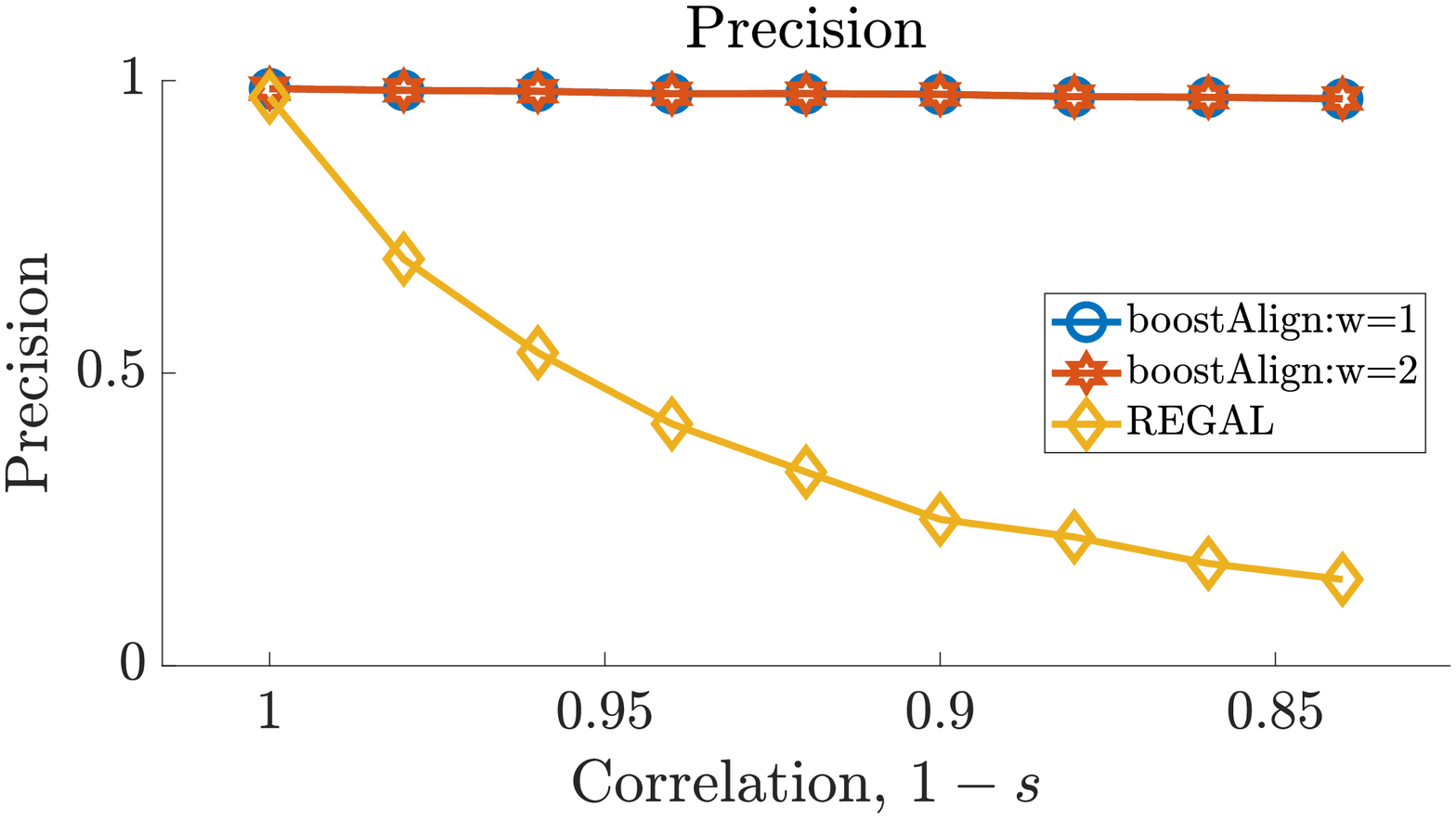}
	\caption{GEMSEC Facebook-Artists\protect\footnotemark}
	\label{subfig:gemsec}
	\end{subfigure}
	\quad
	\begin{subfigure}{0.3\linewidth}
	\includegraphics[width=\linewidth]{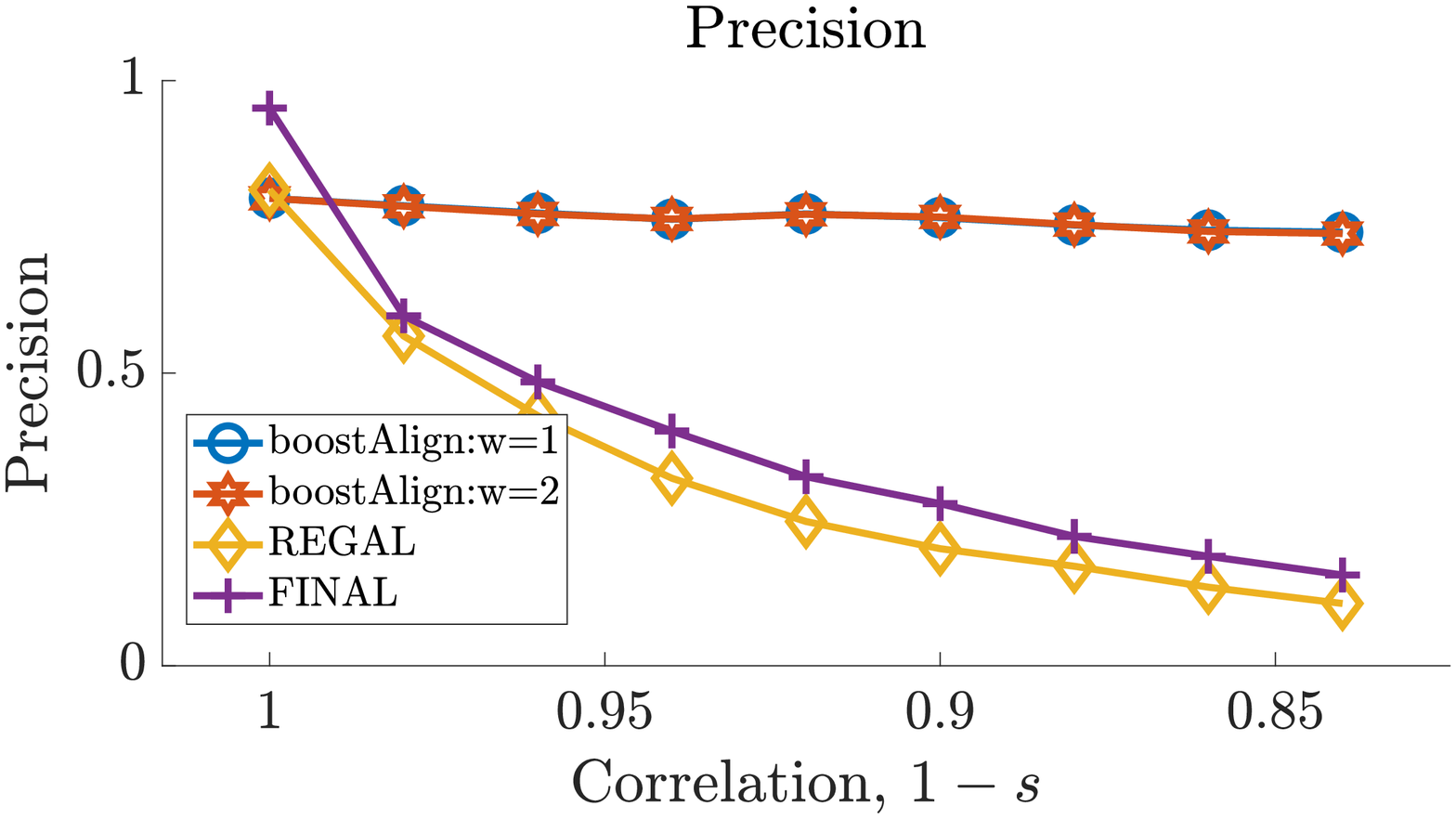}
	\caption{arXiv Astrophysics}
	\label{subfig:astro}
	\end{subfigure}
	\caption{Performance of network alignment methods with varying edge correlation levels. $\alg$ (in dark blue) achieves consistently higher edge correctness than its competitors. }
	\label{fig:comparison}
\end{figure*}

As discussed in Section~\ref{subsec:algone}, the process of seed selection requires the parameters $k$ and $w$ to be chosen. In Figure~\ref{fig:gem_params} we do a sweep of the parameters with $k\in\{20,30,\ldots,100\}$ and $w \in\{1,2,3\}$ on the GEMSEC Facebook-Artists network~\cite{BR-RD-RS-CS:18} to study their influence on the performance of $\alg$. As suggested in Section~\ref{sec:alg}, we fix $f = 3/4$, $r = 4$, and use eigenvector centrality to build the initial noisy seed estimate. In all cases, we limit the number of iterations of $\algtwo$ and $\algthree$ to be no more than five. Intuitively, increasing $k$ should allow for the possibility of more correct pairs to be included, and a larger $w$ increases the probability of placing correct pairs in the initial seed estimate. This intuition is supported by our empirical results; however, increasing these parameters is not free since a larger proportion of incorrect (or noisy) pairs are included in the seed set, as shown in Figure~\ref{subfig:gem_pct}. In a situation with no prior or side information about the network, our empirical results suggest that $k = O(\log n)$, where $n = \min\{|\V_1|,|\V_2|\}$, and $w = 2$ are good choices. Even when the networks to be aligned exhibit a moderate correlation of $60\%$, as shown in Figure~\ref{subfig:gem_prec}, choosing $k = 20$ and $w = 2$ yields near-perfect precision and similarly high recall, suggesting $\alg$ was able to correctly percolate throughout the networks.

We can see the percolation behavior in Figures~\ref{subfig:gem_prec} and~\ref{subfig:gem_rec}; specifically, the efficacy of repeatedly re-running $\algtwo$ and $\algthree$ to boost the alignment performance. In these figures, we observe thresholding behavior; in particular, the matching will either succeed in percolating throughout the networks, achieving high precision and recall, or it will fail and perform poorly. This empirical result is similar to the percolation phenomenon in Erd{\"os}-R{\'e}nyi graphs~\cite{LY-MG:13}, as well as empirical studies on real-world networks~\cite{TS-LC:17}. The main reason for performing the boosting step in $\alg$ is to encourage the success of this percolation. By seeding $\algtwo$ with the previous matching that did not spread well, we may enable the percolation to succeed after further iterations have taken place. In practice this effect induces a much larger and more accurate network alignment. As Figure~\ref{subfig:gem_prec} shows, we may achieve high-quality alignments even in networks which are not very correlated. Another practical consideration is that larger values of $k$ and $w$ tend to induce higher runtimes, as larger noisy seed estimates takes longer to spread in the $\algtwo$ subroutine. This effect can clearly be seen in Figure~\ref{subfig:ppi_runtime}; notice that the behavior is not exactly monotonic due to the randomness in the way $\algtwo$ breaks ties, which in some cases may result in poor seed estimates. However, the largest impact on running time is from the boosting rounds, as each additional iteration of $\algtwo$ and $\algthree$ takes several minutes on large networks.

\subsection{Correlated Networks}\label{subsec:corr}

In order to provide access to the ground-truth node correspondences, we run numerical experiments on correlated networks generated by randomly sampling the edges of a given arbitrary graph $\Gcal$. In particular, we select each edge in $\Gcal$ with a probability $1-s$, independently of other samples. Performing this sampling twice and shuffling the labels of the resulting graphs we obtain two graphs, $\Gcal_1$ and $\Gcal_2$, to be aligned. Since the ground-truth information about node correspondences is available to us, we may measure the accuracy of the alignment $\alg$ produces. Moreover, by tuning the edge dropout probability $s$, we may test our algorithm on pairs of networks with different levels of correlation. Hence, we can measure the performance of $\alg$ as a function of the level of correlation of the networks to be aligned.

\footnotetext{FINAL's MATLAB implementation ran out of memory when attempting to align the Facebook-Artists network.}


We compare two versions of our algorithm, varying the window size $w$ against 2 existing network alignment methods: (i) REGAL~\cite{MH-HS:18} and (ii) FINAL~\cite{SZ-HT:16}. REGAL constructs a node embedding using a low-rank representation and then matches greedily using a fast approximate algorithm. On the other hand,  FINAL is an attributed fast alignment algorithm that extends IsoRank \cite{RS-JX:07}, which in the unattributed case performs a random-walk based fixed-point algorithm to create an embedding based off of an initial similarity matrix $H$. For REGAL we use the default parameters suggested by the authors, but for FINAL we adjust the prior-alignment matrix $H$ to resemble our initialization strategy. We find that initializing $H$ as a sparse matrix with a 1 in pairs of nodes that are within $w$ positions in their respective centrality rankings considerably outperforms the suggested \textit{degree-similarity} initialization.
The results for aligning two (generated) correlated graphs from three large-scale benchmark networks, at varying correlation levels, are presented in Figure~\ref{fig:comparison}. Notably, the PPI networks for \textit{E. Colli} and \textit{C. Jejuni} \cite{YL-KN-AP:15} have been used as a real application for testing alignment algorithms~\cite{VS-TM:14,EK-HH:16,NMD-KB-NP:17,MH-HS:18}.  GEMSEC Facebook-Artists~\cite{BR-RD-RS-CS:18} was used as a representative of a large real-world social network. Lastly, we tested on the benchmark network of collaborations in arXiv for Astrophysics \cite{JL-JK:07}.

\begin{figure}[th]
	\centering
	\includegraphics[width=0.95\linewidth]{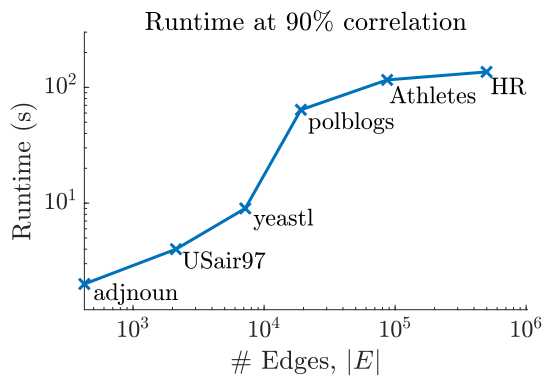}
	\caption{Runtime of $\alg$ on multiple networks varying in number of edges.}
	\label{fig:runtime}
\end{figure}

To observe how the runtime of $\alg$ is effected by the number of edges in the networks to align, we performed several experiments using publicly available datasets of varying size. In each case, we generated two $90\%$ correlated networks using the process described earlier. A plot of the runtime of $\alg$ on each of these networks is given in Figure~\ref{fig:runtime}. The networks include: a word-noun adjacency graph (adjnoun)~\cite{MN:06}; connections between US airports (USair97)~\cite{VB-AM:06}; a yeast PPI network (yeastl)~\cite{VB-AM:06}; a graph of hyperlinks between political blogs (polblogs)~\cite{LA-NG:05}; a network of athletes' pages from Facebook (Athletes)~\cite{BR-RD-RS-CS:18}; and the relationships of Hungarian users of the music streaming service Deezer (HR)~\cite{BR-RD-RS-CS:18}. The properties of all networks used for numerical experiments are shown in Table~\ref{tab:gem_prop}. All experiments were performed on a quad-core Intel Core i7 at 2.2GHz with 16GB of RAM.

\begin{table}[!hb]
\caption{Properties of all networks used in numerical experiments.}
\label{tab:gem_prop}
\begin{center}
\begin{tabular}{l c c c c}
Data set & $|\V| $ & $|\E|$ & Avg.Deg. & Max.Deg. \\
\hline
\hline
\textit{C. jejuni} & $3,294$ & $19,643$ & $11.93$ & $699$ \\

\textit{E. coli} & $1,290$ & $11,100$ & $17.21$ & $154$ \\

Facebook-Artists & $50,515$ & $819,306$ & $32.44$ & $1,292$ \\

arXiv Astrophysics & $18,772$ & $198,110$ & $21.11$ & $354$  \\

Adjective-Noun & $112$ & $425$ & $7.58$ & $49$  \\

US Air & $332$ & $2,126$ & $12.80$ & $139$  \\

Yeast PPI & $2,284$ & $6,646$ & $5.81$ & $64$  \\

Political Blogs & $1,224$ & $19,087$ & $31.18$ & $468$  \\

Facebook-Athletes & $13,866$ & $86,858$ & $12.53$ & $468$ \\

Deezer-HR & $54,573$ & $498,202$ & $18.26$ & $420$ \\
\hline

\end{tabular}
\end{center}
\end{table}

\subsection{Protein-Protein Interaction Networks}\label{subsec:ppi}
In the case of protein-protein interaction (PPI) networks, nodes correspond to proteins, and edges are placed between them if they participate in interactions together. We may not know the ground-truth matching, but we are looking for proteins that perform similar functions across species. We will consider the PPI networks of the bacteria species \textit{Campylobacter jejuni} (\textit{C. jejuni}) and \textit{Escherichia Coli} (\textit{E. coli}) from the HitPredict Database~\cite{YL-KN-AP:15}, which have been used as a benchmark by other algorithms such as MI-GRAAL~\cite{OK-NP:11} and GHOST~\cite{RP-CK:12}. Using $\alg$, we achieve an Edge Correctness of $32\%$ and ICS score of $35\%$, which is a significant increase over both GHOST and MI-GRAAL. 

\begin{figure*}[ht]
	\centering
	\begin{subfigure}[th]{0.3\linewidth}
	\includegraphics[width=\linewidth]{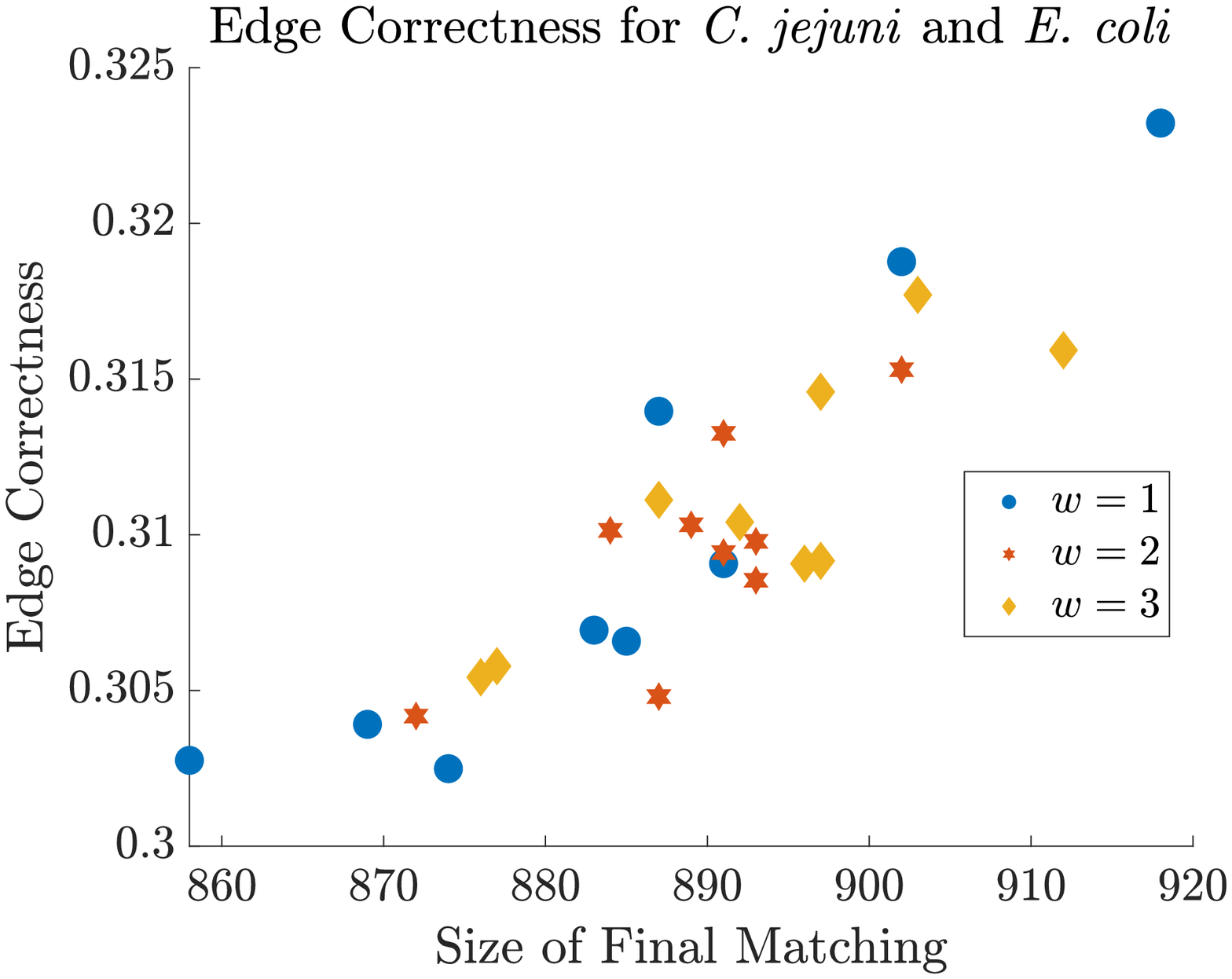}
	\caption{Edge Correctness}
	\label{subfig:ppi_ec}
	\end{subfigure}
	\quad
	\begin{subfigure}[th]{0.3\linewidth}
	\includegraphics[width=\linewidth]{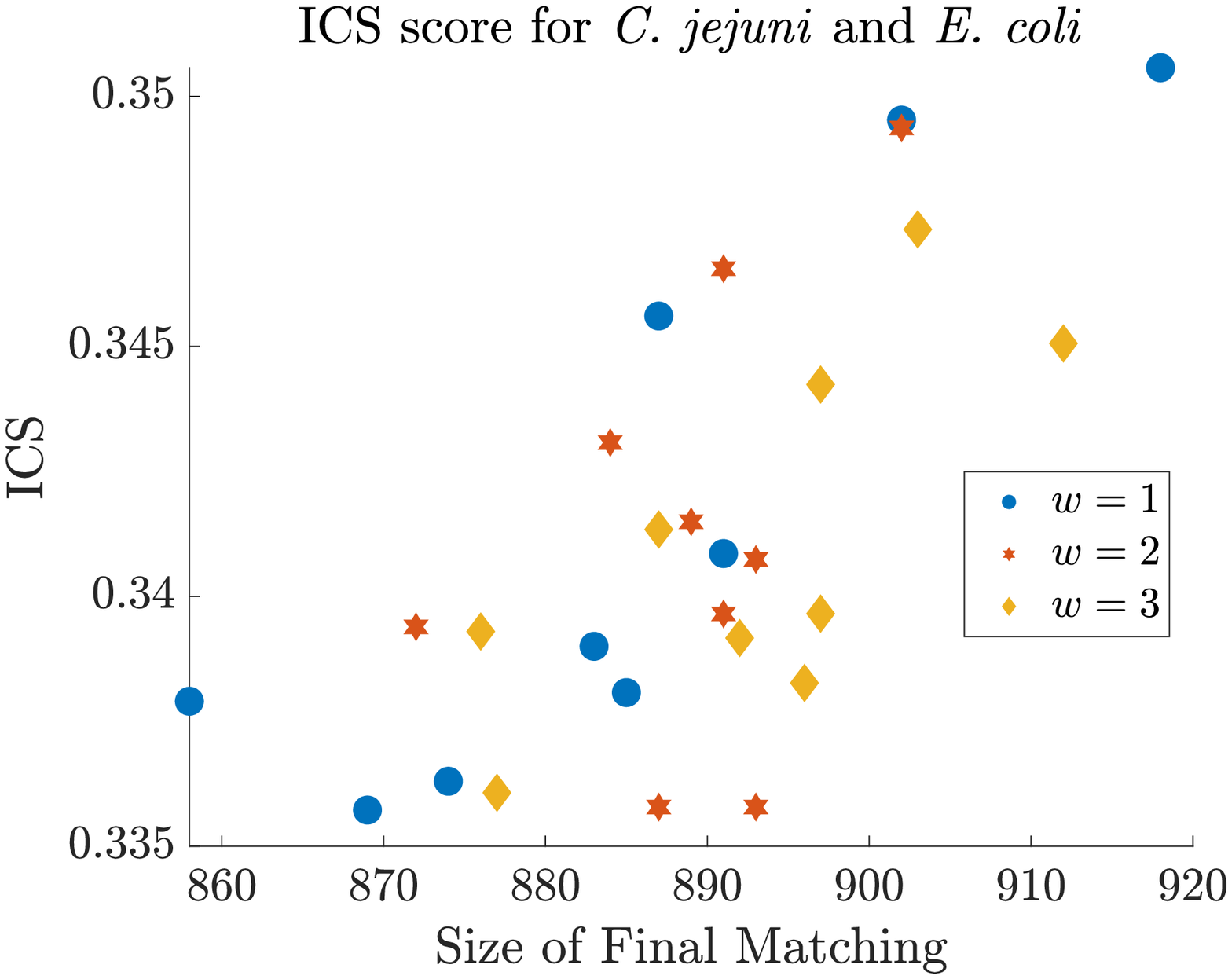}
	\caption{ICS score}
	\label{subfig:ppi_ics}
	\end{subfigure}
	\quad
	\begin{subfigure}[th]{0.3\linewidth}
	\includegraphics[width=\linewidth]{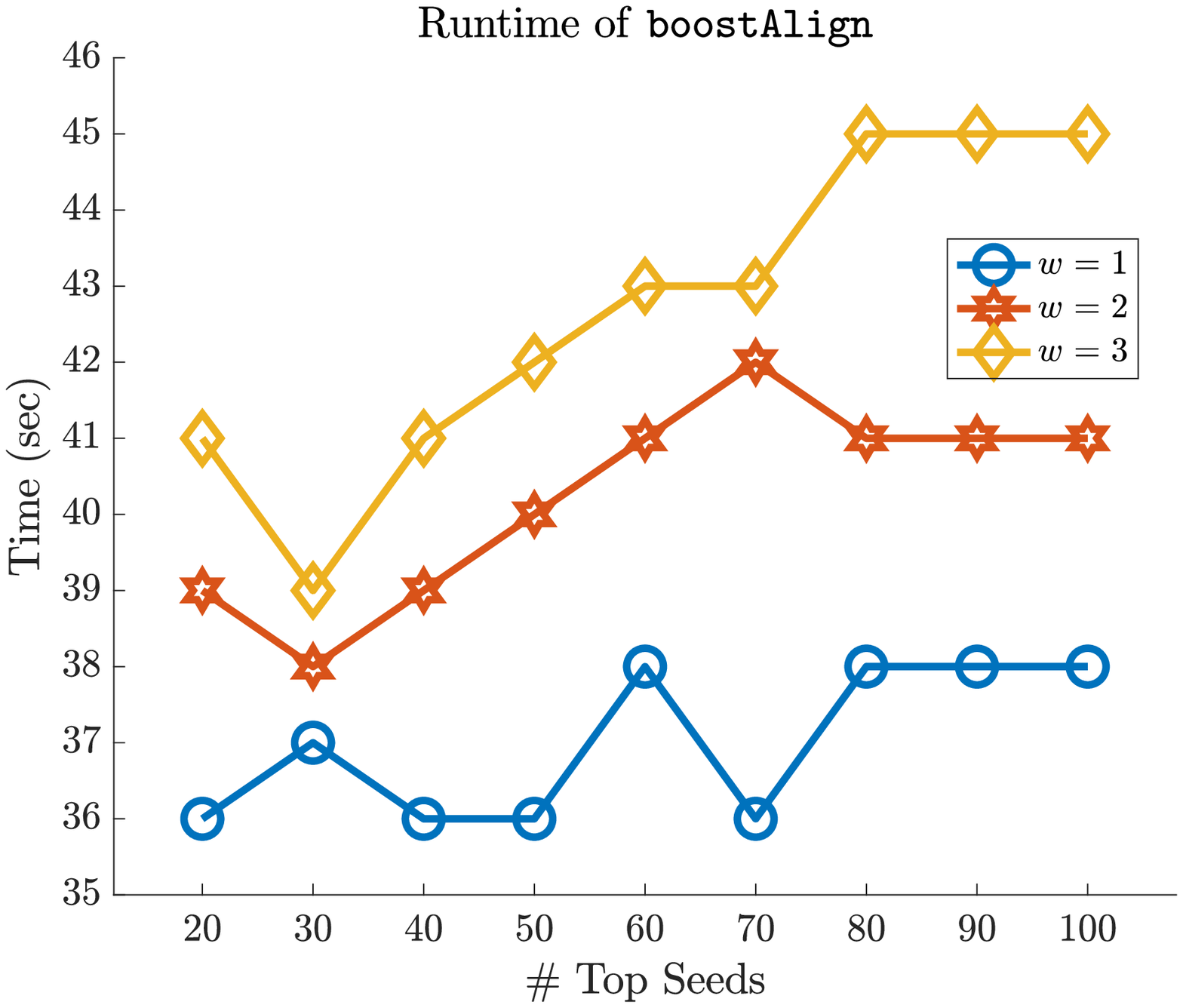}
	\caption{Runtime (sec)}
	\label{subfig:ppi_runtime}
	\end{subfigure}
	\caption{Performance of $\alg$ on \textit{C. jejuni} and \textit{E. coli} PPI networks.}
	\label{fig:ppi}
\end{figure*}

Figure~\ref{fig:ppi} summarizes the results of running $\alg$ on the PPI networks. In these experiments, we fix $f = 3/4$, $r = 4$, use eigenvector centrality to generate the initial seed estimate, and ran a sweep of our parameters with $k\in\{20,30,\ldots,100\}$ and $w \in\{1,2,3\}$. We permitted $\alg$ to run no more than five iterations of $\algtwo$ and $\algthree$.  Again we observe a correlation between the size of the final matching and the quality of the alignment, both as measured by Edge Correctness and ICS score. The runtime of $\alg$ is lowest for $w = 1$ and, interestingly, the best alignment in terms of Edge Correctness and ICS score is for $w=1$, with $k = 90$. These results illustrate that $\alg$ is robust even when aligning large-scale real-world networks without any prior information. Moreover, it outperforms most approaches found in the literature in terms of accuracy, while being much more computationally scalable.

\section{Conclusion and Future Work}\label{sec:conclusion}
We have presented a robust algorithm for aligning networks called $\alg$ which outperforms state-of-the-art algorithms, in terms of both precision and edge correctness, on moderately correlated networks. Indeed, $\alg$ is the first scalable algorithm to exhibit such high precision without using prior information. Unlike most algorithms, $\alg$ uses no seeds or side information. Instead, a noisy seed set estimate is generated using spectral centrality measures. $\alg$ then uses bootstrap percolation techniques along with a backtracking strategy that allows it to iteratively improve the quality of the alignment. We compare its performance with other algorithms found in the literature on a number of benchmark real-world networks from different sources. As the correlation between the networks to be aligned decreases, $\alg$ remains able to obtain high-quality alignments.

The robustness of $\alg$'s performance is due to the bootstrap percolation framework. Future lines of research will extend this framework to the attributed network alignment problem and to the construction of network embeddings. There is also room for improvement in the initialization strategy; because $\alg$ can handle a very noisy seed set, a simple rule for pairing a small subset of $k$ nodes using eigenvector centrality produced remarkable results. We expect that several variants of existing algorithms can produce even more reliable alignments on a small subset of nodes for seeding purposes. Finally, we think these results motivate several theoretical questions regarding the ranking induced by spectral centrality measures and its apparent robustness to edge deletions.


\bibliographystyle{ieeetr}
\bibliography{./biblio,./MH-bib}

\end{document}